\documentclass[aip,jmp,amsmath,amssymb,reprint]{revtex4-1}

\usepackage{graphicx}
\usepackage{dcolumn}
\usepackage{bm}
\usepackage{color}
\usepackage{soul}
\usepackage[]{lineno}

\def\procspie{Proc.\ SPIE}
\def\nat{Nature}
\def\aap{Astronomy and Astrophysics}
\def\apl{Applied Physics Letters}
\def\jltp{Journal of Low Temperature Physics}

\def\aanda{Astronomy and Astrophysics}
\def\pasp{Publications of the Astronomical Society of the Pacific}
\def\rsi{Review of Scientific Instruments}
\def\ji{Journal of Instrumentation}
\def\ieee{IEEE Transactions on Microwave Theory and Techniques}

\begin{document}



\title[]{A Scalable Cryogenic LED Module for Selectively Illuminating Kinetic Inductance Detector Arrays}

\author{J.\ E.\ Shroyer}
\email{jordan.shroyer@virginia.edu}
\author{M.\ Nelson}
\author{L.\ Walters}
\author{B.\ R.\ Johnson}
\affiliation{Department of Astronomy, University of Virginia, Charlottesville, Virginia 22904, USA 
}

\date{\today}


\begin{abstract}

We present the design and measured performance of an LED module for spatially mapping kinetic inductance detector arrays in the laboratory.
Our novel approach uses a multiplexing (MUX) scheme that only requires seven wires to control 480 red LEDs, and the number of LEDs can be scaled up without adding any additional wires.
This multiplexing approach relies on active surface mount components that can operate at cryogenic temperatures down to 10~K. 
Cryogenic tests in liquid nitrogen and inside our cryostat demonstrate that the MUX circuit works at 77~K and 10~K, respectively.
The LED module presented here is tailored for our millimeter-wave detector modules, but the approach could be adapted for use with other KID-based detector systems.

\end{abstract}


                              
\maketitle


\section{Introduction}
\label{sec:introduction}

Kinetic inductance detectors\cite{day_2003, zmuidzinas_2012, mauskopf_2018, mazin_2021} (KIDs) are planar superconducting resonators used to detect light.
The most common variety of KID has a sensing element designed to absorb photons.
If the energy of an absorbed photon is greater than two times the superconducting energy gap of the material, then the absorbed photon will break one or more Cooper pairs in the device.
This changes the quasiparticle density, the complex impedance of the resonator, and subsequently, the resonant frequency of the KID, 
which is detected as a change in the amplitude and phase of a probe tone driving the resonator.
KIDs are being designed and fabricated for a broad range of wavelengths and applications\cite{shu_2022, glenn_2021, vissers_2020, fyhrie_2020, choi_2020, tang_2020, sayers_2020, walter_2020, basset_2019, mccarrick_2018, johnson_2018, steinbach_2018, barry_2018, redford_2018, mazin_2018, meeker_2018, mcgeehan_2018, austerman_2018, hubmayr_2018, vavagiakis_2018, cataldo_2018, jones_2017, baselmans_2017, ritacco_2017, bueno_2017, barrentine_2016, dober_2016, griffin_2016, szypryt_2016, mccarrick_2014, hubmayr_2014, noroozian_2013, brown_2010, mazin_2006a}.

The KID technology is designed with large arrays in mind.
Each detector in an array has its own resonant frequency, and frequency division multiplexing can be used to read out hundreds or thousands of detectors with a single transmission line\cite{duan_2020, gordon_2016, bourrion_2016, van_rantwijk_2016, mccarrick_2014, mchugh_2012}.
The precise resonant frequency of each detector in the array depends on the capacitance, the geometric inductance, and the kinetic inductance of the device.
The capacitance and geometric inductance are sensitive to fabrication tolerances.
The kinetic inductance is difficult to know before fabrication; resonator designs are commonly based on fiducial values of the kinetic inductance fraction for the chosen material.
Consequently, the actual resonant frequencies of fabricated devices may differ somewhat from the designed resonant frequencies, so the physical location of each resonant frequency in the array can therefore be difficult to forecast.
A measurement is usually needed to determine the physical array address of each KID.

For astrophysical observations, for example, a detector array map is commonly measured by observing a compact far-field source like a planet.
By synchronously recording the detector output and the telescope pointing, it is possible to determine the physical location of each detector in the focal plane of the instrument.
A similar laboratory-based measurement is difficult, so mapping the KID array during detector development can be challenging.


\begin{figure}
\includegraphics[width=0.9\columnwidth]{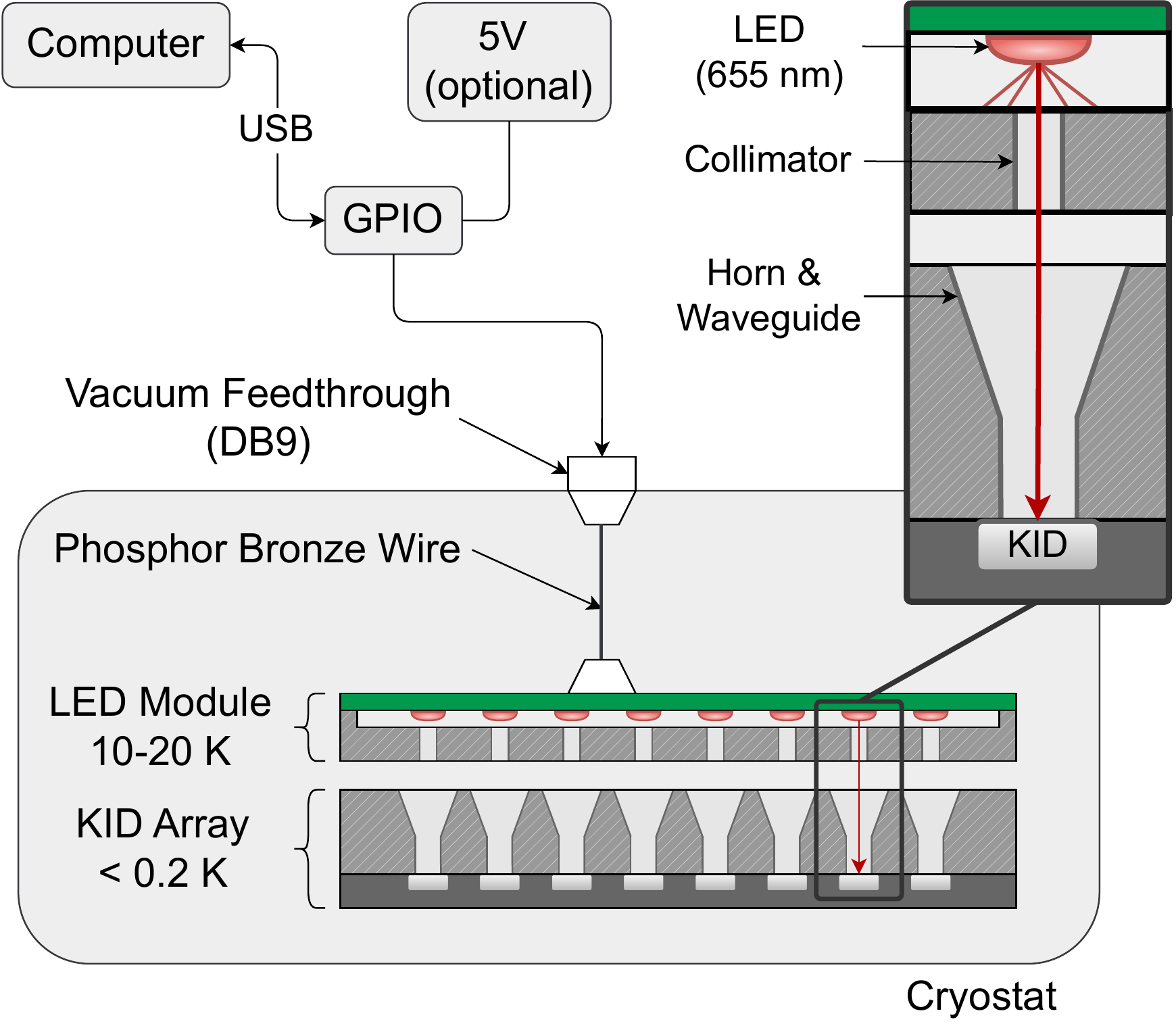}
\caption{
System diagram.
An LED module is mounted in front of the KID array inside the cryostat.
One LED is illuminated at a time and the location of its KID partner is revealed.
As an example, this schematic depicts a horn-coupled KID array.
It is important to note that the LED module is mounted to a different temperature stage in the cryostat, and it does not mechanically touch the KID module.
}
\label{fig:system_diagram}
\end{figure}


\begin{figure*}
\includegraphics[width=0.8\textwidth]{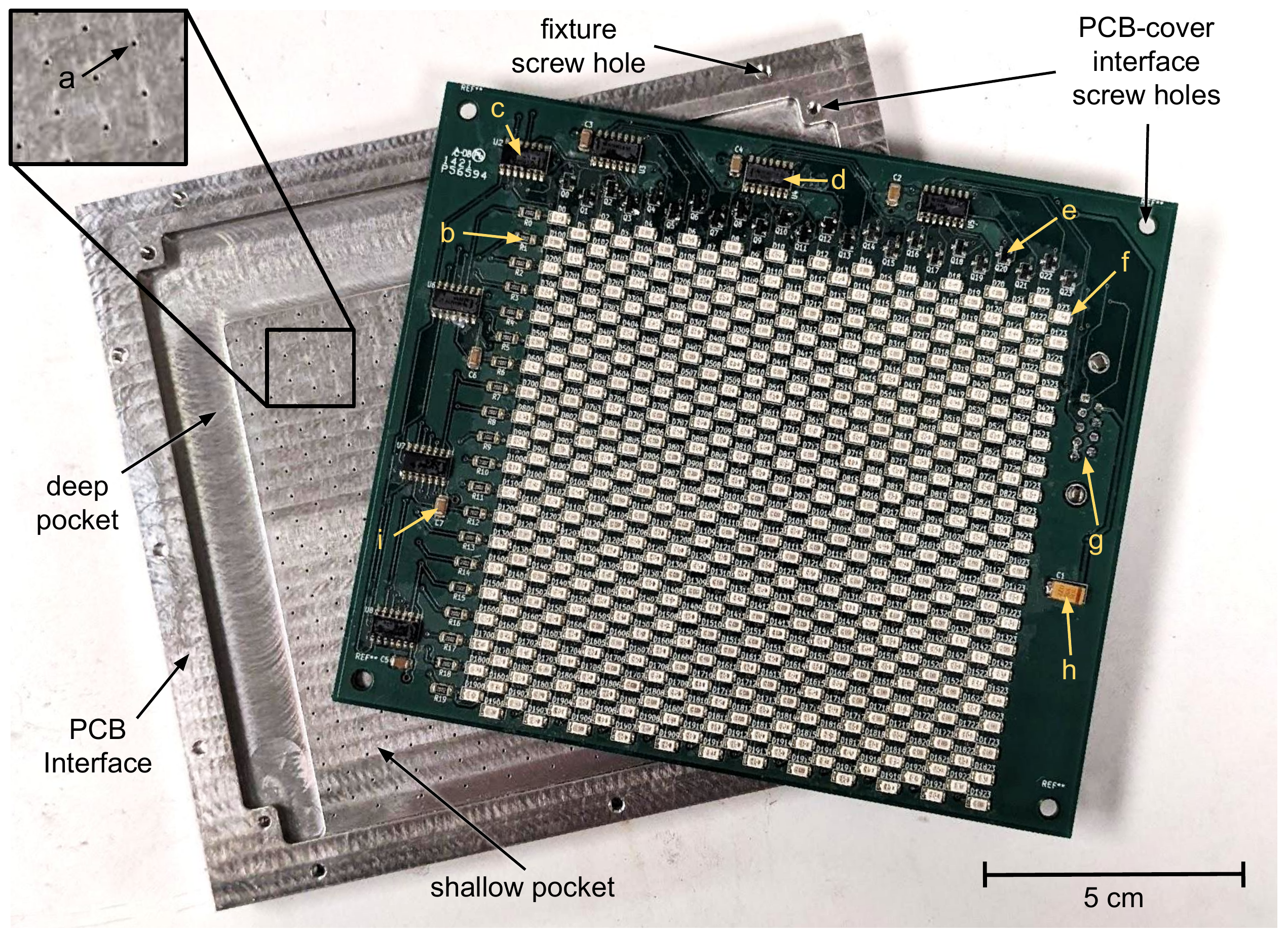}
\caption{
Photograph of the LED array and the aluminum cover.
The key elements of the LED array are labeled: $(a)$ one of the 480 collimators, $(b)$ one of the 20 current-control resistors on each row (R1 in Figure~\ref{fig:led_array_circuit}), $(c)$ all four OR gates, $(d)$ one of the six shift registers, $(e)$ one of the 24 MOSFETs, $(f)$ one of the 480 LEDs, and $(g)$ the back side of the surface-mount DB9 connector.
The yellow capacitors $(h, i)$ that can be seen at various locations on the PCB are used to ensure power stability.
To assemble the module, the LED array board is flipped over to meet the PCB interface on the cover, each LED aligns with one collimator (as shown later in Figure~\ref{fig:LED_detail}), and four screws fasten the LED array to the cover. 
Nine tapped fixture screw holes around the outer edge of the PCB interface can be used to mount the module inside the cryostat.
}
\label{fig:led_array_photo}
\end{figure*}


\begin{figure*}
\includegraphics[width=\textwidth]{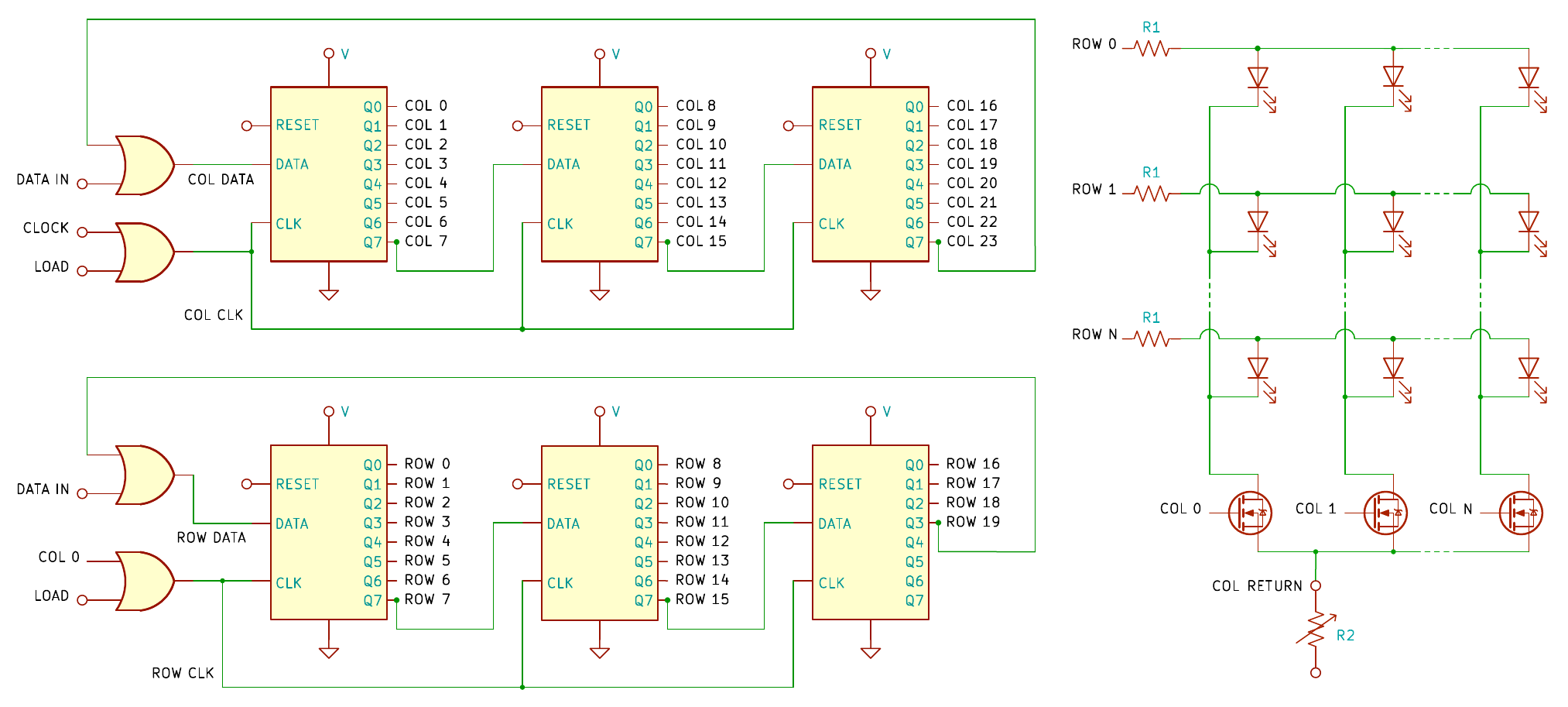}
\caption{
Schematic of the LED array circuit.
A total of seven wires are used to control and power 480 LEDs. 
The yellow rectangles (left) are the six 8-bit shift registers which comprise the two-dimensional multiplexer.
The top three shift registers control the columns in the LED array, while the bottom three shift registers control the rows; to synchronize the two, COL~0 is tied to ROW~CLK (see Section~\ref{sec:circuit_board}). 
The shift register outputs connect to the corresponding rows or columns of the LED array (right).
The ``ROW''  outputs pass through resistors to limit current through the LEDs. 
The  ``COL'' outputs connect to the gates of MOSFET transistors which serve as switches to close the circuit for the selected column.
The ``R2" resistor is a potentiometer mounted outside the cryostat, and it is used to fine-tune the LED brightness.
The open red circles correspond to the points where the seven external signals from the GPIO connect to the LED array (see Section~\ref{sec:control_system_hardware}).
The ground symbol shown (called GROUND in the text) is connected to the GPIO ground and not the cryostat case.
}
\label{fig:led_array_circuit}
\end{figure*}


One laboratory-based approach for spatially mapping the resonators in a KID array is shown in Figure~\ref{fig:system_diagram}.
An array of light emitting diodes (LEDs) is mounted in front of the KID array with one LED above each KID array element.
The LEDs are then turned on one at a time.
The resonant frequency of the KID below the illuminated LED shifts, and the array address of that particular KID corresponds to the address of the illuminated LED.
One technical challenge with this approach is the LED array needs to be mounted in front of the KID array, so the LED array needs to be cooled to, and function at, cryogenic temperatures.

\citet{liu_2017a} developed this kind of technology for a KID array with 126 elements.
Their program focuses on correcting the resonant frequencies of the KIDs.
They first map the KID array with their LED array, and then they lithographically edit the resonators so the actual resonant frequencies match the design\cite{liu_2017b}.
Their LED array design uses 16 DC bias wires to address the LEDs.
The Liu \textit{et al.} design is scalable; in an optimized configuration, their approach uses $m$ wires to address an array of up to $m(m+1)$ LEDs.

We built a similar LED array to test a millimeter-wave detector system composed of horn-coupled KID modules that contain $20 \times 24 = 480$ KID elements.
In this paper, we present our new approach, which uses a multiplexing scheme that only requires seven wires to control $20 \times 24 = 480$ red LEDs, and the number of LEDs can be scaled up without adding any additional wires.
This multiplexing approach relies on active surface mount components that can operate at cryogenic temperatures down to 10~K. 
The LED module we designed and built is described in Section~\ref{sec:methods}.
Section~\ref{sec:results} describes our tests, which demonstrate that the active components work cryogenically.
We know our KIDs\cite{mccarrick_2018, mccarrick_2014} respond to red LED light, so this paper focuses on describing the LED module technology and its development.
The LED module presented here is tailored for our millimeter-wave detector modules, but the approach could be adapted for use with other KID-based detector systems.


\section{Methods}
\label{sec:methods}


\subsection{Requirements and System Overview}
\label{sec:requirements_system_overview}

For our application, there are four primary requirements for the LED module.
First, there needs to be one LED per KID array element, and photons from one LED should illuminate one KID array element without adversely affecting the neighboring KIDs (Requirement~\#1).
For example, stray light inside the KID module from excessive illumination could produce cross-talk in neighboring KIDs, making the measurement result difficult to interpret.
Second, the LED module needs to be mounted in front of the KID array (Requirement~\#2).
This requirement presents a challenge because our KIDs are typically operated below 0.2~K \cite{mccarrick_2018, mccarrick_2014}.
The LED module does not need to operate at this low temperature, too, but it should operate at a temperature that is similar to the temperature of the thermal background loading for which the KIDs are optimized.
For example, millimeter-wave KIDs for astrophysical observations are usually loaded by, and optimized for, unavoidable atmospheric signal, which has a brightness temperature of approximately 10 to 20~K.
If the temperature of the LED module is too hot (significantly above 20~K, for example), then thermal emission from the LED module (not the LED photons themselves) could saturate the KIDs.
Third, the thermal loading on the cryogenic stage that hosts the LED module needs to be within the thermal budget of the cryogenic cooling system (Requirement~\#3).
The LED module can contribute to thermal loading through the electrical control wires from 300~K, or through dissipation in the active electrical components in the module.

Finally, the photons from the LEDs need to be energetic enough to break Cooper pairs (Requirement~\#4).
BCS theory predicts that the superconducting energy gap is $\Delta = 1.764 \, k T_c$, where $k$ is the Boltzmann constant and $T_c$ is the superconducting transition temperature of the material\cite{tinkham_2004}.
Therefore, photons with a frequency greater than the cutoff frequency
\begin{align}
\nu_{\mathrm{cutoff}} = \frac{2 \Delta}{h} \approx 74~[\,\mathrm{GHz}\,] \times \left( \, \frac{T_c}{1~\mathrm{K}} \, \right)
\end{align}
will break Cooper pairs in the KID.
Our KIDs use thin-film aluminum, which has $T_c$ = 1.4~K, so photons with frequencies greater than approximately 100~GHz will be energetic enough to break Cooper pairs in our KIDs~\cite{mccarrick_2018, mccarrick_2014}.
The red photons produced by the LEDs we chose have $\nu$ = 460~THz (more in Section~\ref{sec:circuit_board}).

A schematic of the system design is shown in Figure~\ref{fig:system_diagram}.
The system is composed of the LED module mounted inside the cryostat (Section~\ref{sec:module_design}), the control system mounted outside the cryostat (Section~\ref{sec:control_system}), and the cryogenic wires, which connect the two (Section~\ref{sec:cryo_wiring}).


\subsection{Module Design}
\label{sec:module_design}

The LED module consists of a printed circuit board (PCB) and an aluminum cover, shown in Figure~\ref{fig:led_array_photo}.
The PCB hosts the array of LEDs and the active multiplexing components.
The aluminum cover includes an array of collimators that are used for LED light management, and it serves as a thermal bus and a mechanical interface.


\subsubsection{LED Array}
\label{sec:circuit_board}

The control circuit on the PCB was nominally designed to minimize the number of external signals required to operate an arbitrarily-sized two-dimensional array of LEDs inside a cryostat, thereby minimizing the heat input onto the cryogenic stage through the wiring (see Requirement~\#3).
A schematic of our circuit design is shown in Figure~\ref{fig:led_array_circuit}.
The circuit is a two-dimensional multiplexer (MUX) built from six 8-bit shift registers.
Three of these shift registers are combined to produce one 24-bit shift register that controls the column-select functionality.
The remaining three are combined to produce a 20-bit shift register that controls the row-select functionality.
This MUX design allows for row/column scanning, individual LED illumination, and blinking.
The design is scalable because additional 8-bit shift registers can be added as needed for larger arrays.

Seven external wires are passed into the cryostat to control the LED array via the MUX circuit: V, GROUND, DATA~IN, RESET, LOAD, CLOCK, and COL~RETURN. 
The connection points for these wires can be seen in the schematic in Figure~\ref{fig:led_array_circuit}.
A constant 5~V is used to power the shift registers (V). 
GROUND is the control system ground and not the cryostat case (more in Section~\ref{sec:control_system_hardware}). 
DATA~IN is the signal that will be passed around the shift registers in the MUX (5~V during initialization, otherwise 0~V). 
The 8-bit shift registers we chose\footnote{Texas Instruments P/N: CD74AC164M96 (This device has SOIC packaging.)} each have two serial data inputs, either of which can be used. 
We chose to tie them together for this application.

Initialization is accomplished in three steps.
First, the output pins (labeled Q0 to Q7 in Figure~\ref{fig:led_array_circuit}) of all the 8-bit shift registers are reset to 0~V by toggling RESET\footnote{The reset pin on the shift registers is actually $\overline{\mathrm{RESET}}$, so the input is toggled from  5~V to 0~V and back to 5~V. RESET is always set to 5~V during operation.}.
Then, the DATA~IN signal is set to 5~V and passed to the shift registers through an OR gate\footnote{Texas Instruments P/N: CD74AC32M96}.
Finally, a LOAD  signal (5~V) is used to pass the output of that OR gate into the first bit of both the row-select and the column-select shift registers. 
As an example, if DATA~IN is set to 5~V, then the LOAD signal will set COL~0 and ROW~0 to 5~V, which will in turn illuminate the LED at the (0,0) address in the LED array.

The OR gates at the CLK inputs of the row-select and column-select shift registers allow us to use a single clock signal to advance both the columns and the rows, rather than introducing separate wires to clock the row and column registers independently. 
In the column shift register, LOAD shares an OR gate with an external clock signal (CLOCK), which is used to advance the DATA~IN voltage through the column register after initialization.
The column registers are shifted on the positive edge of the clock pulses.
In the row register, LOAD shares an OR gate with the first bit (COL~0) of the column shift register.
This is the point of contact between the column and the row shift registers; COL~0 is tied to ROW~CLK in Figure~\ref{fig:led_array_circuit}.
After a full loop through the column-select shift register is completed, the row increments by one.
The OR gates at the DATA inputs of the row-select and column-select shift registers serve to roll around each full register. 
When the CLOCK signal is periodically toggled, the illuminated LED continuously scans through the array from left to right, top to bottom (in the orientation shown in Figure~\ref{fig:led_array_photo}).
The LOAD signal is nominally for initialization but it can also be used to advance the illuminated LED, simultaneously advancing both the row and the column shift registers.

The actual on/off column switching is done with MOSFETs\footnote{Diodes Incorporated P/N: BSN20-7} on the low-voltage side of the LEDs.
The 5~V outputs of the column-select shift registers are connected to the MOSFET gates.
Each column ends with a MOSFET, and all MOSFETs are connected through a common return line (COL RETURN).
The LEDs are driven by the 5~V outputs of the row-select shift registers dropped through current-limiting resistors (R1 in Figure~\ref{fig:led_array_circuit}).
We chose to use LEDs with GaAlAs chips\footnote{Lumex P/N: SML-LX23SRC-TR} that have a peak wavelength of 655~nm (see Requirement~\#4).
The current through the LEDs can be controlled by changing the on-board R1 resistors, and fine-tuned with the R2 potentiometer on the COL~RETURN line, which is mounted outside the cryostat.
For R1, we started with 20~k$\Omega$ thick film chip resistors\footnote{Vishay Dale P/N: CRCW120620K0FKEAC} but ultimately chose to use 75~k$\Omega$ metal film resistors\footnote{KOA Speer Electronics, Inc.\ P/N: RN73H2ATTD1003F}.
The resistance of the metal film resistors appears to be more stable at cryogenic temperatures (more in Section~\ref{sec:discussion}).

To keep the LED array cold, electrical power dissipation needs to be sufficiently small (see Requirement~\#3).
Power dissipation on our PCB is very low because the active components are complementary metal oxide semiconductor (CMOS) devices, which use field-effect transistors.
CMOS components similar to the components we chose have been demonstrated to work at cryogenic temperatures down to approximately 4~K\cite{buchanan_2012}.
At 300~K with R1 set to 75~k$\Omega$ and R2 set to 0~$\Omega$, our measurements show the LED module draws 50~$\mu$A of current with all of the LEDs off, and 100~$\mu$A with just one LED on.
These measurements suggest the MUX circuit alone dissipates 250~$\mu$W, and 50~$\mu$A flows through the LED and the R1 resistor when the LED is on.
Therefore, in this configuration, the measured forward voltage of the LED is 1.3~V, and the LED dissipates approximately 60~$\mu$W of power either as heat or light.
Our cryogenic system has a cooling power of 0.7~W, so this heat load is acceptable for our application.
If all 60~$\mu$W of power in the LED is converted to photons, then the emerging signal would be too bright for our KIDs.
Consequently, we added the R2 potentiometer to the circuit and collimators to the cover (more in Section~\ref{sec:aluminum_cover}) to attenuate and control the brightness of the LED signal reaching the KIDs.
Other attenuation strategies that we may use in the future are discussed in Section~\ref{sec:discussion}.

The PCB is a two-layer FR-5 board that is 1.6~mm thick with no ground plane.
Prototype PCBs were made by PCBFabEx, and the final PCBs were made by Advanced Circuits.
The LED array is laid out in a hexagonal packing arrangement with a 4.8~mm pitch.
The footprint of each surface-mount LED component is 2$\times$3~mm, and they are 1.3~mm tall with a clear epoxy lens.
The surface mount components were soldered onto the PCB in our laboratory with a reflow oven\footnote{The reflow oven was made by Whizoo, and it uses the Controleo3 controller.}.
Capacitors were added to the circuit to ensure power stability.
A single 10~$\mu$F tantalum capacitor\footnote{KEMET P/N: T491C106K016AT} connects the 5~V power pin on the surface-mount DB9 to ground, and a 1~$\mu$F surface mount multi-layer ceramic chip capacitor\footnote{KEMET P/N: C1206C105K4RACTM (SMD MLCC)} connects the V pin of each 8-bit shift register to ground.


\begin{figure}
\includegraphics[width=0.8\columnwidth]{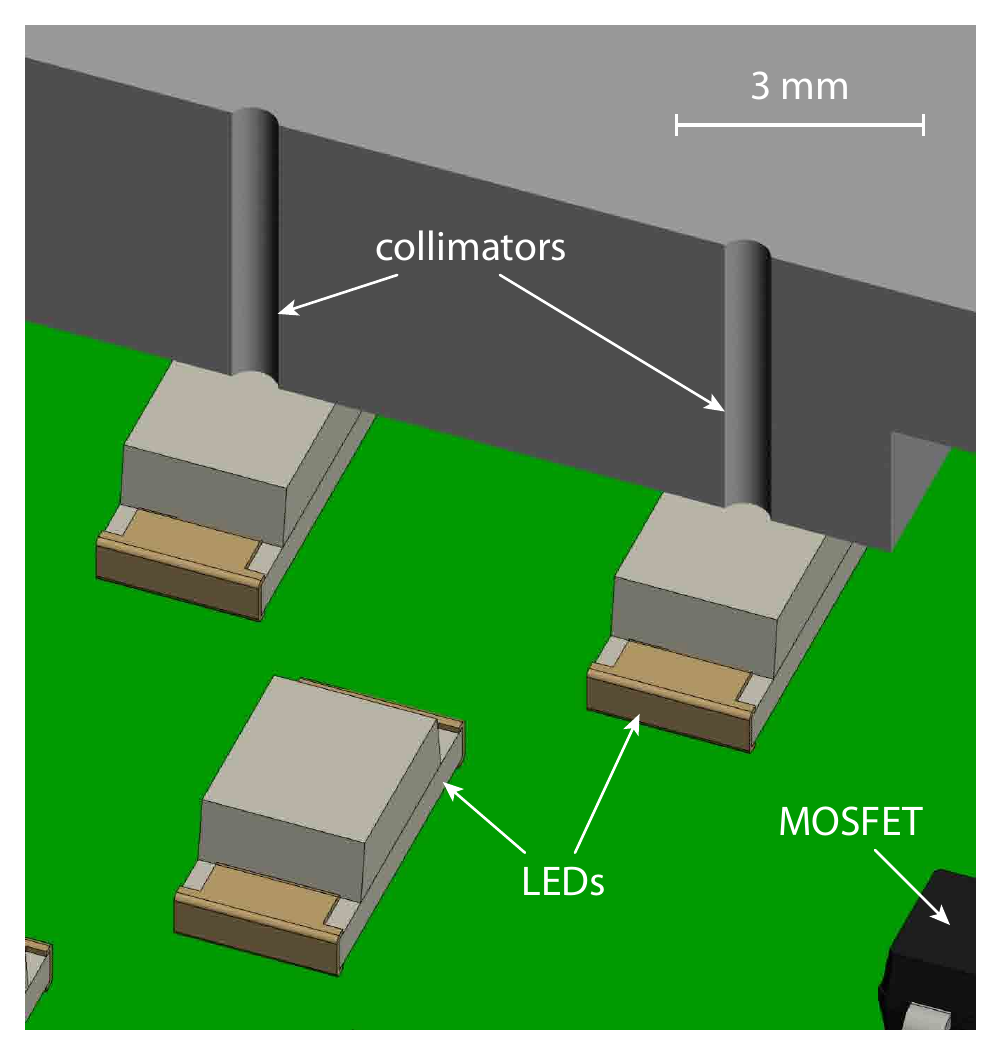}
\caption{
Model of the LED module showing the collimators.
The aluminum cover is shown in cross section to highlight the way the collimators align with the LEDs.
There is a 180~$\mu$m gap between the top of the LEDs and the entrance aperture of each collimator.
The dimensions of the parts in the model are at 300~K.
Consequently, the collimators are not centered over the LEDs before cooldown.
During cryogenic operation, the aluminum thermally contracts more than the PCB, and the collimators become centered over the LEDs.
}
\label{fig:LED_detail}
\end{figure}


\subsubsection{Cover with Collimators}
\label{sec:aluminum_cover}

We do not want light from one LED illuminating multiple KID array elements (see Requirement~\#1).
The LED viewing angle is 120~deg, which makes optical cross-talk possible if the bare LED array is used.
To help manage the emitted photons, a custom aluminum cover was made \footnote{The cover was fabricated by ProtoLabs}.
This cover was machined from a piece of 6061-T6 aluminum that is 155\,$\times$\,137\,$\times$\,4.78~mm, and it is mounted over the LED array.
A pocket was milled into the surface of the aluminum cover facing the LED array to make room for the various surface-mount components, and a hole 330~$\mu$m long and 50~$\mu$m in diameter was machined through the cover above each individual LED.
These holes serve as collimators for the LED photons.
The collimators can be seen in the photograph in Figure~\ref{fig:led_array_photo}, and Figure~\ref{fig:LED_detail} shows a cross-sectional view of a three-dimensional CAD model of the aluminum cover highlighting the collimator/LED alignment.
The pocket is 1.48~mm deep around the LED array.
This depth sets the LEDs 180~$\mu$m away from the entrance aperture of the collimators.
Away from the LEDs, the pocket is 3.01~mm deep.
This deeper pocket ensures there is enough space for the other surface mount components, some of which are taller than the LEDs.
The aforementioned dimensions are for 300~K.

The dimensions of the features machined into the aluminum cover -- including the collimator locations -- were selected with differential thermal contraction in mind.
During cryogenic operation, the collimator pitch and the LED pitch need to be the same, and the cover and the PCB need to be laterally aligned so the LEDs and the collimators line up properly.
We also do not want the PCB to break from induced stress.
The integrated coefficient of thermal expansion (CTE) from 100 to 293~K is $\Delta \ell/\ell = 4.15 \times 10^{-3}$ and $\Delta \ell/\ell =  2.79 \times 10^{-3}$ for aluminum and epoxy fiberglass, respectively; from 4 to 100~K, the CTE is $\Delta \ell/\ell = 0.47 \times 10^{-3}$ for both materials\cite{weisend_1998}.
Therefore, the differential CTE we assumed is $1.36 \times 10^{-3}$ from 10~K to 293~K.
As a consequence, the aluminum cover is somewhat oversized at 300~K.
The PCB and the aluminum cover are laterally aligned using four interface screws, one in each corner. The interface screw holes on the PCB are 3~mm in diameter.
On the aluminum cover, one of the interface screw holes is tapped for an M3$\times$0.5 screw, while the other interface screw holes are tapped for M2.5$\times$0.45 screws.
The M3 screw serves as the point of reference for the differential thermal contraction because there is a close fit between the screw and the hole.
The holes are oversized for the other three screws, allowing the necessary movement.
Spring washers were used on all four screws to ensure good thermal contact between the PCB and the cover.
The aluminum cover also has nine additional tapped fixture screw holes around the perimeter that were added to allow the assembled LED module to be mounted to fixtures inside the cryostat.


\subsection{Cryogenic Wiring}
\label{sec:cryo_wiring}

Signals travel from the control system outside the cryostat (see Section~\ref{sec:control_system_hardware}) to a board-mounted DB9 connector\footnote{Amphenol P/N: LD09P24A4GV00LF} on the LED module through a DB9 vacuum feedthrough~\footnote{CeramTec NW40KF DSUB vacuum feedthrough P/N: 18605-01-KF}.
The wiring is shown schematically in Figure~\ref{fig:system_diagram} and the surface-mount connector can be seen in Figure~\ref{fig:led_array_photo}.
For the seven wires inside the cryostat, we used 32~AWG phosphor bronze (CuSnP alloy) wire with polyimide insulation\footnote{We used 32 AWG phosphor bronze wire from Lake Shore Cryotronics: single lead (P/N: WSL-32) for COL~RETURN, Duo-Twist (P/N: WDT-32) for V and GROUND, and a Quad-Twist (P/N: WQT-32) for DATA~IN, CLOCK, LOAD, and RESET.}.
Our calculations show that the heat load from 300 to 3~K through 1 meter of 32~AWG phosphor bronze wire is approximately 1~mW, so approximately 7~mW for the seven-wire harness\footnote{We used the thermal conductivity values for phosphor bronze provided by Lake Shore Cryotronics.}.
This heat load is acceptable for our testing system, which has 0.7~W of cooling power at 3~K, so we made the wiring harness 1~meter long and did not heat sink the wires between the vacuum feedthrough at 300~K and the LED module.


\begin{figure}
\centering
\includegraphics[width=\columnwidth]{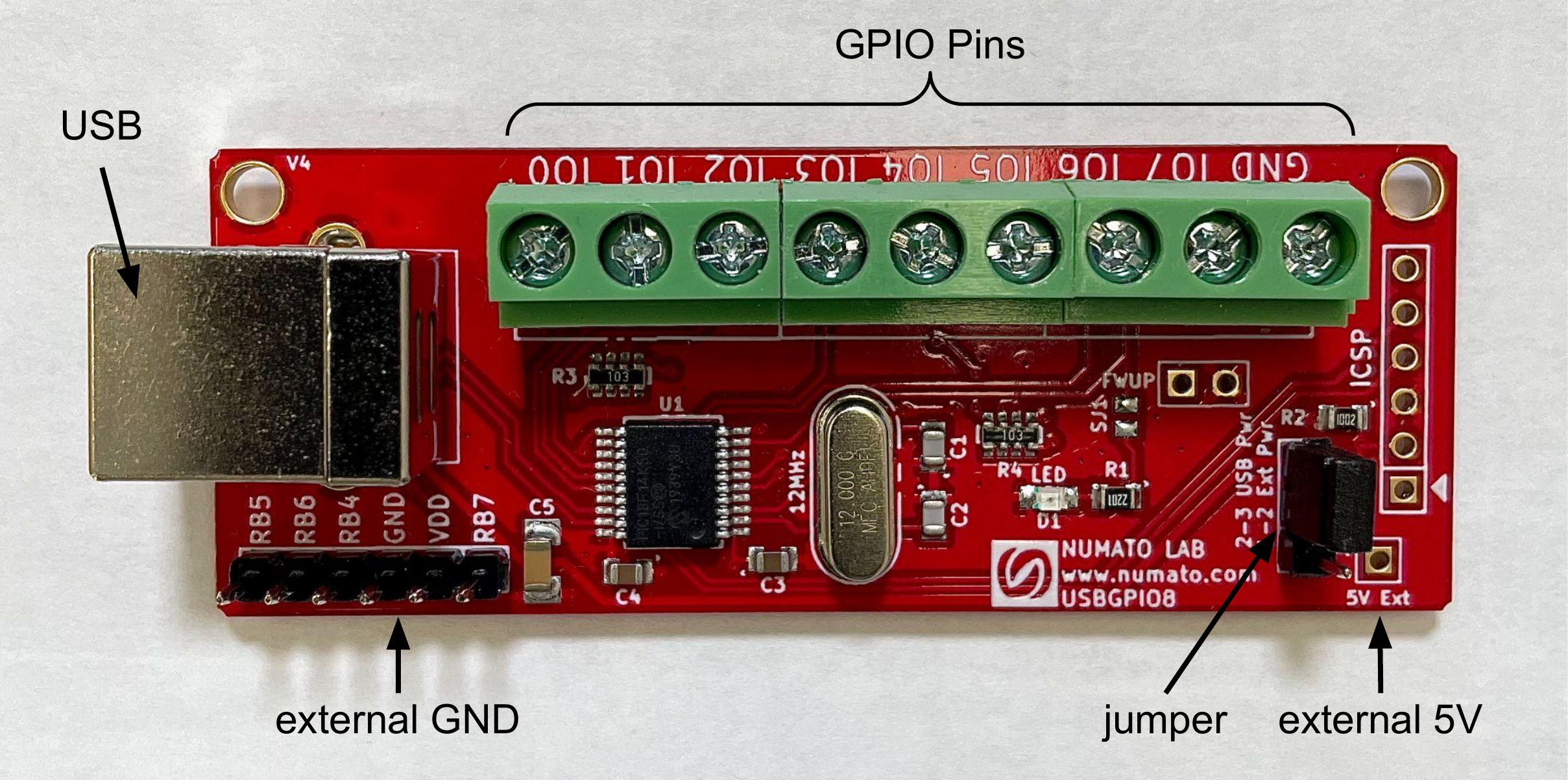}
\caption{
A photograph of the GPIO that is mounted inside the control electronics box (see Figure~\ref{fig:electronics_enclosure}).
The USB receptacle on the left connects to the control computer.
The GPIO pins across the top connect to the LED module (see Section~\ref{sec:circuit_board}).
The external 5~V connection can be seen in the bottom right-hand corner next to a jumper, which is used to switch between USB power and the optional external 5~V power.
}
\label{fig:gpio}
\end{figure}


\subsection{Control System}
\label{sec:control_system}


\subsubsection{Hardware}
\label{sec:control_system_hardware}

The control system consists of an 8-channel general purpose input/output (GPIO) module\footnote{Numato Labs P/N: GP80001} (see Figure~\ref{fig:gpio}) and a control computer running a custom software application written in Java (more in Section~\ref{sec:control_system_software}).
The control computer communicates with the GPIO module using standard USB protocols.
The GPIO module features eight 5~V IOs, and a ninth pin for ground.
Our design uses six pins in digital mode\footnote{Six of the IOs can be used as 0 to 5~V digital or analog inputs, the other two are exclusively digital.} and the ground pin to control the circuit with the seven external signals described in Section~\ref{sec:module_design}.
During operation, the V pin that powers the shift registers is always set to high (5~V).

\begin{figure}
\centering
\includegraphics[width=.9\columnwidth]{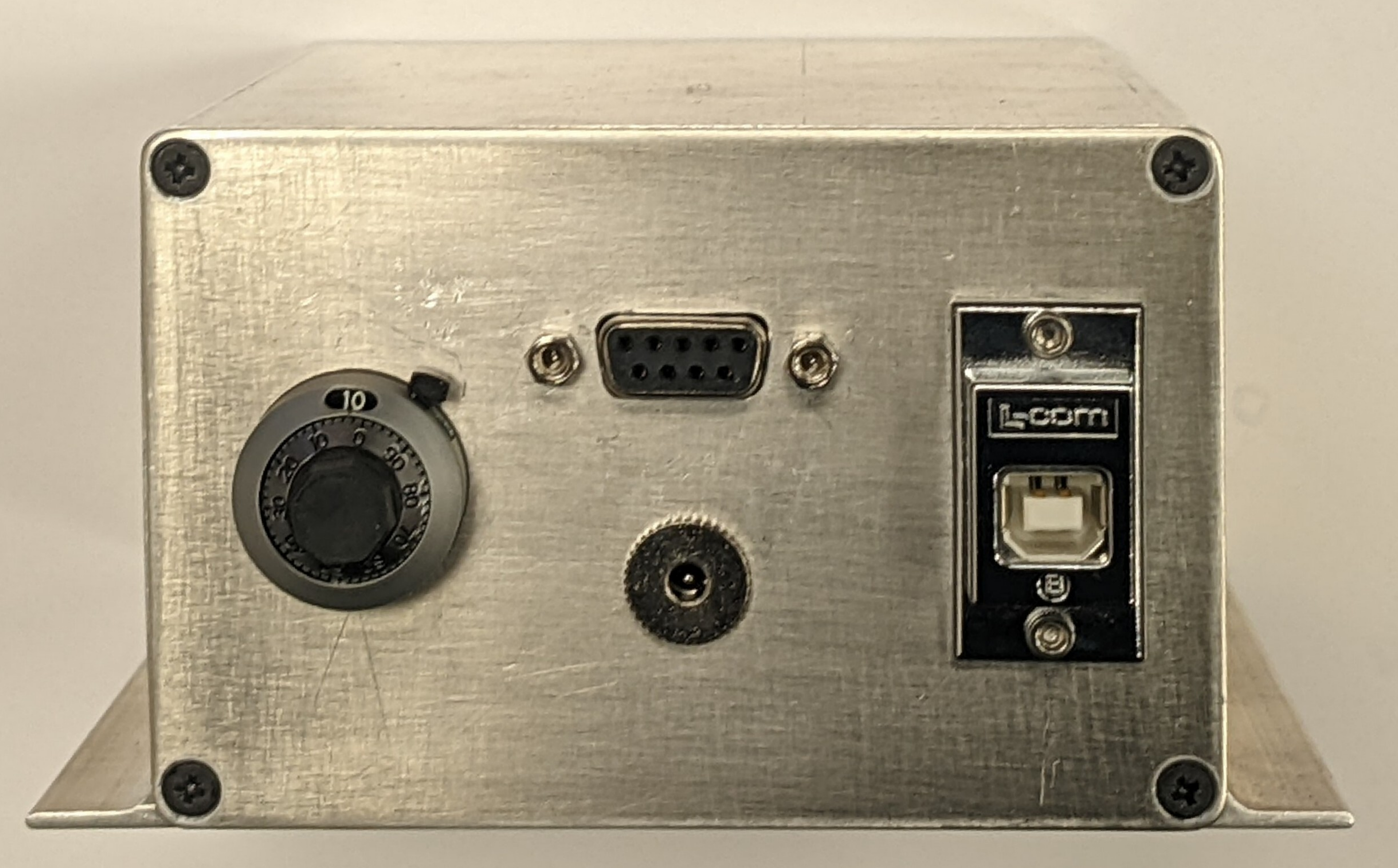}
\caption{
A photograph of the control electronics box.
The GPIO and the potentiometer that controls the LED brightness (R2 in Figure~\ref{fig:led_array_circuit}) are mounted inside an extruded aluminum enclosure.
The potentiometer dial is on the left. 
The panel mount DB9 connector in the top center connects to the cryostat, and it carries signals from the GPIO to the LED module.
The panel mount USB type B connector on the right connects the GPIO to the control computer (not shown).
The panel mount barrel connector in the bottom center is the optional 5~V power jack.
}
\label{fig:electronics_enclosure}
\end{figure}

The GPIO and the R2 potentiometer are mounted in an extruded aluminum enclosure\footnote{McMaster-Carr P/N: 3686K11 and P/N: 3686K15} (see Figure~\ref{fig:electronics_enclosure}).
Commands are sent from the control computer to the GPIO module via USB.
A 9-pin serial cable carries signals from the electronics box to the vacuum feedthrough and into the cryostat.
The control electronics box uses a panel mount USB type B coupler~\footnote{L-Com P/N: ECF504-12BBS} to communicate with the control computer and a panel mount DB9 connector for the outgoing signals\footnote{NorComp Inc.\ P/N: 171-009-203L021 (socket), P/N: 171-009-103L021 (pin), P/N: 160-000-011R031 (jackscrew)}.

The LED array is nominally powered by the control computer via the USB connection.
Alternatively, an external 5~V power supply can be used if power stability is a concern.
If the external 5~V power is used, then a jumper on the GPIO needs to be set.
The corresponding panel mount power connector is a barrel connector jack\footnote{Tensility International Corp.\ P/N: 54-00061} (1.35~mm ID, 3.50~mm OD).


\subsubsection{Software}
\label{sec:control_system_software}

The user controls the LED module with custom Java software (hereafter app) that is designed to run on Windows, Linux, and Mac.
Commands can be sent from the control computer to the GPIO using either a graphical user interface (GUI) or the command line.

The array is initialized automatically when the app starts up, and it can be re-initialized manually at any time. 
With the app, the user can advance a single row or column, or illuminate an LED at an any chosen array position.
The LEDs blink by toggling the value of COL~RETURN.
If COL~RETURN is 0~V, then the LED is on. 
If COL~RETURN is 5~V, then the LED is off.
The user can modify the scan rate and blink rate, which modifies the delay between the commands sent from the computer to the GPIO.

We added a 3~ms delay in the software after each write to the GPIO to ensure that commands are not sent from the computer more quickly than the GPIO can process them. 
We did not try to optimize the delay time; we simply chose a value that reliably prevented commands from piling up. 
This delay time was acceptable for our application. 
Advancing the column requires six writes to the GPIO, so the entire 480-LED array can be scanned in less than ten seconds.
To access an arbitrary element, the user toggles LOAD to advance diagonally through the LED array, stopping ahead of the desired array element. %
Then, CLOCK is toggled to advance the column to the desired array element. 
Advancing diagonally also takes six writes to the GPIO. 
For an $n$-pixel square LED array, the random access time is $(2\times18$~ms$)\times \sqrt{n}$. 
For example, in our nearly square array, the final LED is at (row, col)~=~(19, 23). 
We toggle LOAD 19 times to get to (18, 18), and CLOCK is toggled 4~+~23 times to reach (19, 23). 
This takes (18 ms)$\times$19~+~(18 ms)$\times 27 = 828$~ms. 
Any arbitrary LED in our array can be accessed in less than a second.
If we scale to a square ten kilopixel array, for example, a scan will take approximately 3~min, and turning on an arbitrary LED will take approximately 4~s. 
The blink rate is variable and is also limited by the same 3~ms software delay. 
It takes 9~ms to turn the LED off, and another 9~ms to turn it back on. 
Therefore the highest frequency blink rate is 1/18~ms~=~56~Hz.

The control software compensates for two trade-offs of our design.
First, the wiring scheme does not preclude the illumination of multiple LEDs at once.
The app tracks the array state to ensure only one LED is lit at a time.
Second, to access an arbitrary array position, the shift registers must be clocked until the desired LED is lit.
The app enables random access by automatically and efficiently advancing the shift registers when the user enters a target row and column.
The user can also choose to save a log of the board status which records the row, column, blink rate, and scan speed at any time during operation.


\begin{figure}
\includegraphics[width=.9\columnwidth]{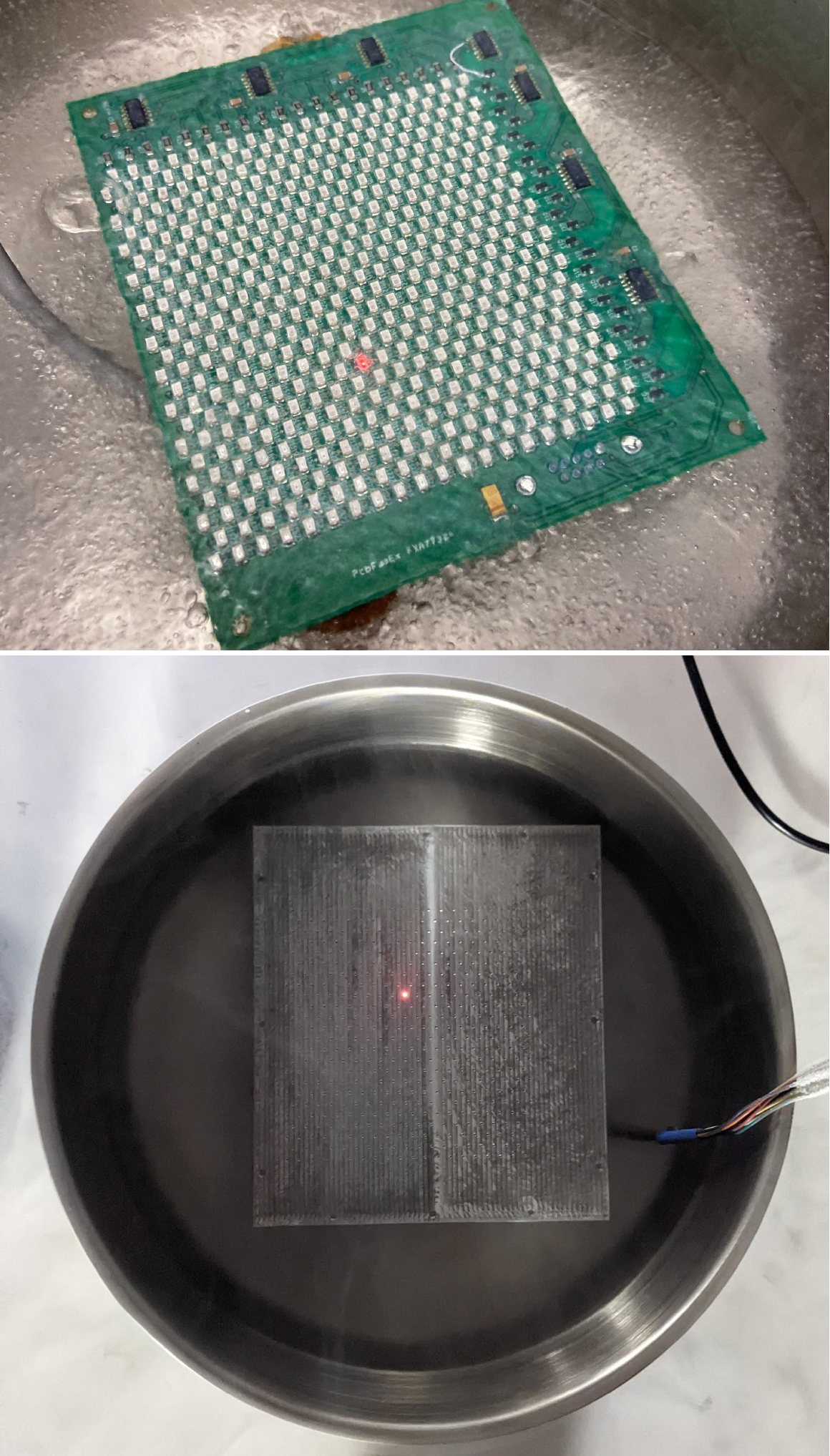}
\caption{
Top: Photograph of LED array submerged in liquid nitrogen.
For this test the LED array was conformal coated to prevent electrical shorts.
Bottom: Photograph of the assembled LED module (LED array with the cover attached), submerged in liquid nitrogen. 
This test shows the LED module functions as designed.
Note that in both photographs one LED is illuminated, showing the board is functioning.
}
\label{fig:ln2-test}
\end{figure}


\begin{figure*}
\centering
\includegraphics[height=0.27\textheight]{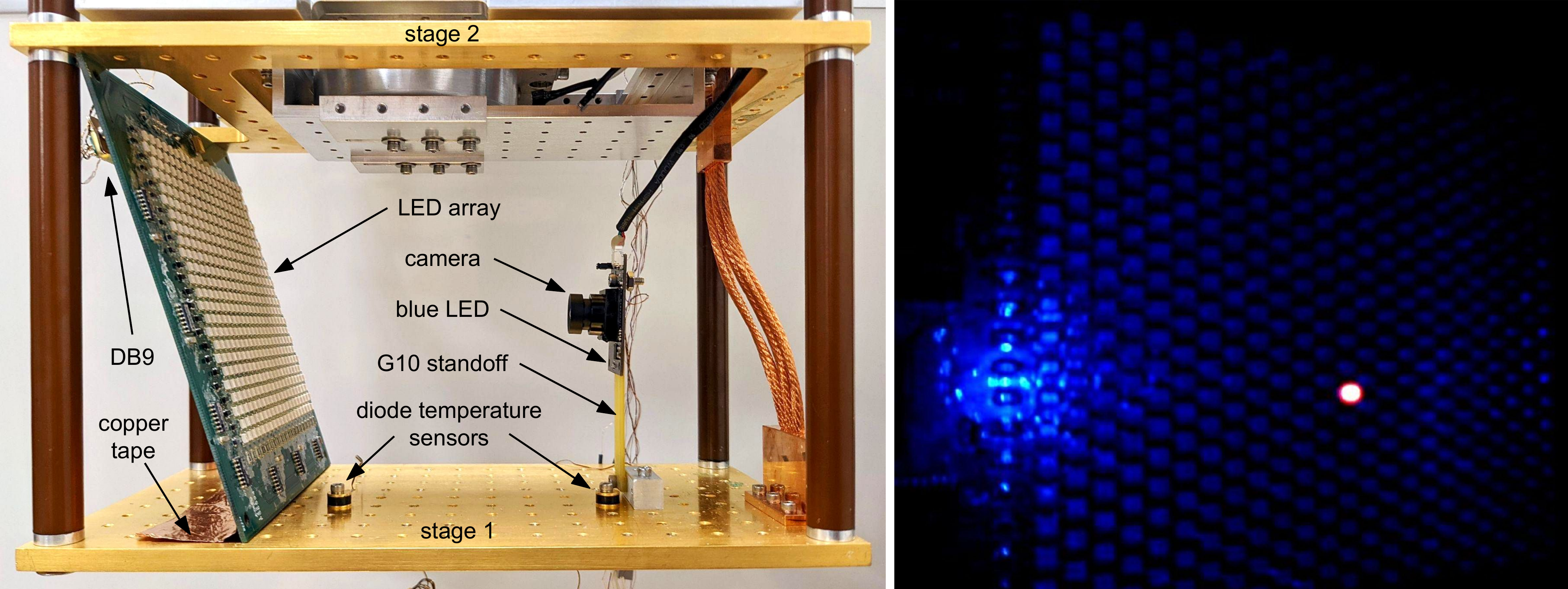}
\caption{
Left: Experimental setup for cryostat testing. 
The LED array is on the left, with a strip of copper tape along the back to enhance thermal contact. 
The DB9 with the cryogenic wires described in Section~\ref{sec:cryo_wiring} can be seen in the upper left. 
The array is monitored with a camera mounted on a G10 standoff for thermal isolation.
The array was illuminated by a blue LED that is mounted on the camera to verify that the camera was operational.
The testing stage temperature is monitored with two diode sensors, one near the array and the other near the camera.
Stage 1 and stage 2 are both in thermal contact with the 3~K stage of the PTC.
Right: Photograph of LED array operating inside the cryostat.
In this photo, the LED array was cooled to 20~K.
Blue light from the LED on the camera can be seen reflecting off of the LED array.
}
\label{fig:cryostat_setup}
\end{figure*}


\section{Results}
\label{sec:results}


\subsection{Liquid Nitrogen Test}
\label{subsec:ln2test}
To quickly test basic PCB functionality and survivability of the surface mount components, the LED array was submerged in a liquid nitrogen (LN2) bath.
For these tests, we used a double-wall, thermally-insulating stainless steel bowl to hold the LN2 (see Figure~\ref{fig:ln2-test}), and the LED array was supported by short phenolic tubes to prevent the bowl from creating electrical shorts.
To prevent moisture from affecting the LED array, we painted all circuit components on the PCB with a conformal coating\footnote{Silicone Modified Conformal Coating \#422B from MG Chemicals} before submersion.
The board was set to scan continuously as it was submerged and while it cooled to equilibrium at 77~K.
Scanning allowed us to verify that the MUX circuit was performing as expected, and it showed whether the LEDs were still functioning.
During these 77~K tests, the board functioned nominally.
However, we did notice in our first test that the LED brightness increased, which indicated that the resistance of the thick film R1 resistor decreased with temperature. 
Consequently, we increased the resistance of the R1 resistors from 20~k$\Omega$ to 75~k$\Omega$.
After testing the circuit board alone and verifying it worked, we attached the cover and tested the fully assembled LED module at 77~K.
We found that the LEDs were aligned under the collimators as designed, the LED array functioned nominally, and there were no mechanical or electrical failures from differential thermal contraction.


\subsection{Cryostat Test}
\label{subsec:cryostat_test}

We use a DRC-102 ADR Cryostat System made by STAR Cryoelectronics for our KID testing\cite{johnson_2018, mccarrick_2018, mccarrick_2014}.
This cryogenic system is capable of cooling to sub-kelvin temperatures using a Cryomech PT407 pulse tube cooler (PTC) together with a two-stage adiabatic demagnetization refrigerator (ADR)\cite{bennett_2012}.
The PTC provides 40~K and 3~K stages, and the ADR nominally cools testing stage~1 and stage~2, shown in Figure~\ref{fig:cryostat_setup}, to $\sim$0.1~K and $\sim$1~K, respectively.
For our LED module test, we reconfigured this system and used it in a non-standard way leaving both stage~1 and stage~2 thermally connected to the 3~K stage of the PTC via a mechanical heat switch (not shown in Figure~\ref{fig:cryostat_setup}).
Therefore, for this study without any heat input, stage~1 and stage~2 are both nominally 3~K.

For our tests, we mounted the LED array on stage~1.
A strip of copper tape was applied across the back edge of the PCB to help ensure the LED array was thermally connected to the stage.
The LED array functionality was monitored with a live video feed from a CMOS camera\footnote{Leopard Imaging P/N: LI-OV5640-USB-72} also mounted on stage~1 inside the cryostat on a G10 standoff\cite{johnson_2017}.
The G10 standoff allowed the camera to self heat, so it operated at a temperature that was higher than the stage~1 temperature\footnote{We did not measure the camera temperature.}.
Since the testing stage in the cryostat nominally cools to 3~K, we initially planned to add heaters to elevate the temperature of the LED array to approximately 10~K.
However, we found the parasitic heat load on stage~1 from the camera serendipitously heated the stage to the correct temperature, so we did not add any heaters.   
The temperature of stage~1 was monitored using diode sensors mounted next to the LED array and the camera.
As before with the LN2 tests, the LED array was set to scan continuously as the temperature dropped from room temperature to 10~K.
Our tests showed that the surface mount components functioned properly when the stage temperature was 10~K because the scan progressed as expected and the array responded to control signals.

We did four separate cryostat tests.
In the first cryostat test we used thick film resistors for R1, R2 was not installed, and the LED failure rate was 3\% (13 out of 480 LEDs).
We suspected this failure rate was due to excessive current through the LEDs as the R1 resistance decreased with temperature.
We tested this hypothesis by increasing R1 and R2 in subsequent cryostat tests.
After increasing R1 and R2, fewer LEDs failed.
We also tried different MOSFETs.
We never changed the shift registers, the OR gates, or the LEDs.
In the final cryostat test we used the original MOSFETs with metal film resistors (see Section~\ref{sec:circuit_board}), and the failure rate was 0.2\% (one out of 480 LEDs).
Since we were focused on getting the module running, we did not study the resistors or the MOSFETs in more detail because this module configuration was meeting our requirements.


\section{Discussion}
\label{sec:discussion}

In the final cryostat test, the LEDs became very dim at 10~K, and many were not visible with the camera\footnote{It is worth noting that the LEDs may have actually been on, but the short, fixed exposure time of the camera prevented us from detecting the faint LED signals.}.
However, when we warmed the LED array back up to $\sim$20~K, all of the LEDs were visible, so we know the LEDs did not permanently fail.
Something caused the LEDs to dim or switch off when the stage temperature was between 20~K and 10~K.
One explanation is the resistance of the metal film resistor we used increases as the temperature decreases below 20~K.
An alternate explanation is that the LEDs are dimming around 10~K due to a change in the properties of the LED. 
This hypothesis requires further investigation.
This dimming effect is not a problem for our application because we can operate the LED module at 20~K if necessary.

The precise temperature of the LED array is of little practical importance for our application because the aluminum cover will be mounted between the LED array and the KID array.
Aluminum has high thermal conductivity so dissipated heat in the LED array should be transported away easily (see Requirement~\#3), and aluminum has low emissivity, so the anticipated thermal emission should be acceptable (see Requirement~\#2).
Rather than conducting a detailed cryogenic study of CMOS components, our cryogenic tests to date have focused on demonstrating that this is a functional tool at cryogenic temperatures.
Therefore, the temperatures listed for the LED array in this paper should be taken as lower limits until a more detailed study is completed.

Going forward, we will test the LED module with a KID module.
Our future testing program will focus on controlling the LED brightness so we satisfy Requirement~\#1.
We plan to tune the LED emission to the right power level for the KIDs by adjusting the R1 resistances, the R2 potentiometer, or both.
If the power on the KIDs is still excessive, we will decrease the collimator diameter or fill the collimators with a lossy dielectric.
If reflection off the inside of the aluminum collimator causes stray light emission, we will add an absorber inside the module and/or black anodize the aluminum.
A future paper will describe the LED module being demonstrated with KIDs.


\section{Conclusion}
\label{sec:conclusion}

We developed an LED module to map the resonator locations in KID arrays.
This module utilizes a multiplexing scheme that is straightforwardly scalable to larger arrays of LEDs without additional wires. 
We demonstrated the system for an array of 480 LEDs.
Cryogenic tests in a liquid nitrogen bath and inside our cryostat demonstrate that the chosen surface mount components work at 77~K and 10~K, respectively.
Future work will include testing a KID module with the LED module we developed.


\bibliography{references}


\end{document}